\documentclass[11pt,twoside]{article}
\usepackage{asp2004}
\usepackage{epsf}
\usepackage{psfig}
\usepackage{lscape}
\begin{document}
\title{Proper Motions as an Underutilized Tool for
Estimating Distances and Ages for Nearby, Young Stars}
\author{Eric E. Mamajek}
\affil{Harvard-Smithsonian Center for Astrophysics, 
60 Garden St., MS-42, Cambridge, MA, 02138 USA}

\begin{abstract}
The recent availability of accurate proper motion catalogs for millions
of stars on the sky (e.g. Tycho-2, UCAC) can 
benefit projects for which age estimates are needed for
stars that are plausibly young ($<$100 Myr) and within a 
few hundred pc of the Sun.
Here I summarize how accurate proper motions have been useful in 
(1) identifying new, nearby, post-T Tauri star
populations, and (2) estimating distances to young field stars
which are lacking trigonometric parallax measurements. The
later enables the calculation of 
stellar luminosities and isochronal ages -- two critical
quantities for investigations of the
evolution of star/planet/disk systems. 
\end{abstract}

\vspace*{-0.75cm}
\section{Motivation}
One of the broad aims of modern astrophysical research is to 
understand how stars and planetary systems form and evolve.
This topic is one of the major research themes driving 
the construction of the next generation 
of ground-based telescopes (e.g. GMT, GSMT), as well as 
future space observatories (e.g. SIM, JWST, TPF-C). 
To study the evolution of star/planet/disk systems, we
require accurate stellar ages. 
As remarkable as the astronomical discoveries
of circumstellar phenomena have been over the past decade 
(e.g. exoplanets, brown dwarfs, debris disks), all too 
often the age of the host star under study is constrained
to an accuracy of tens of percent (at best!). Improved
means of robustly estimating stellar ages are sorely needed.
Stauffer (2004) said it best: ``{\it It would be a waste of 
all the wonderful IR data if the age estimates from
[Spitzer] stellar targets were left as an after-thought.}''

Ages estimates for late-type stars (cooler than mid-F) are 
usually determined through comparing stellar properties 
like activity, rotation, and Li abundance, to stars with 
calibrated ages (usually stars in well-studied clusters). 
Independently, one can estimate an isochronal age through
placing the star's HR diagram position on theoretical evolutionary 
tracks.
{\it Trigonometric parallaxes} enable the 
calculation of luminosity, and hence, isochronal ages.
{\it Proper motions}, which are much easier to measure
than trigonometric parallaxes, 
have been largely underutilized in the 
quest for estimating stellar ages. In the instance
where we have an empirical velocity model for
a sample of objects (say, proposed members of a star cluster),
we can statistically test whether a star is consistent
with co-moving with that group, and use the star's proper
motion to calculate a {\it secular parallax}.

Here I briefly summarize how
proper motions have been useful in finding new, nearby
young stars, and in estimating ages for previously known
specimens that lack trigonometric parallaxes.
The results stem from ancillary science done 
for the Spitzer Legacy program ``Formation and
Evolution of Planetary Systems'' 
(FEPS\footnote{http://feps.as.arizona.edu/}
; Meyer et al. 2004).
These results are presented in more detail in
\citet{Mamajek02,Mamajek03,Mamajek04,Mamajek04phd},
Mamajek (in prep.) and Hillenbrand et al. (in prep.).

\section{Using Proper Motions to Find Post-T Tauri Stars}

Proper motions can aid in discovering low-mass members of 
nearby stellar associations (i.e. ``post-T Tauri stars''). 
\citet{Mamajek02} conducted a survey for such stars over 
hundreds of square degrees in the
two nearest OB subgroups to the Sun (Lower Centaurus-Crux
(LCC) and Upper Centaurus-Lupus (UCL); ages $\simeq$ 15 Myr)
using both proper motion and X-ray selection. 
The memberships of these groups among early-type stars is 
excellent thanks to Hipparcos satellite 
astrometry \citep{deZeeuw99}. Assuming that the
groups had a normal initial mass function, it was very plausible
that a large, untapped reservoir of post-T Tauri stars could
be found co-moving with the high-mass membership. 
Using both proper motion and X-ray selection, \citet{Mamajek02} 
identified $\sim$100 G/K-type pre-MS members of UCL and LCC. Proper
motion selection yielded $\sim$3$\times 10^3$ Tycho stars with 
motions consistent with membership 
\citep[][and priv. comm.]{Hoogerwerf00}, and
$\sim10^3$ ROSAT All-Sky Survey Bright Source 
Catalog \citep[RASS-BSC][]{Voges99} sources blanket the region. 
Selection by X-ray emission and proper motions yielded a 93\%
hit-rate in identifying Li-rich pre-MS G/K-type stars, whereas
using proper motions and parallaxes \citep[][where available]{deZeeuw99},
yielded a hit rate of 73\%. Selecting by proper motions
{\it alone} has proved to be rather inefficient at selecting
faint members of OB associations \citep{Preibisch98},
however when used in conjunction with other datasets 
(i.e. X-rays, color-magnitude diagrams), proper
motions are excellent at rejecting interlopers.

\section{Using Proper Motions to Estimate Distances and Ages}

Proper motions are useful for assessing membership of stars
to clusters and associations, as well as estimating
secular parallax distances for those stars selected as 
members. The selection of cluster members, and
calculation of improved distances to the individual
members using proper motions,
has been successfully attempted for several
nearby clusters and associations 
\citep[e.g.][]{deBruijne99,Madsen02,Mamajek02}.
These individual distances are useful for reducing
the scatter in HRD positions, and searching for
substructure in associations \citep{Mamajek02}. 

Outside of clusters and associations, secular parallaxes
can be used for any kinematic group where the velocity dispersion
is much smaller than the mean heliocentric space motion.
This is the situation for young ($<$100 Myr-old) field
stars in the solar neighborhood, which are moving
with respect to the Sun at $\sim$20 km/s, but have a 1D
velocity dispersion of $\sim$5 km/s. Theoretically one should
be able to calculate distances to tens of percent accuracy to any star
whose tangential velocity is $>$few$\times$ the 1D velocity dispersion
of the group to which it belongs. 
Numerous active, Li-rich, 
late-type stars (presumably young) have been discovered in recent years, but have
either no trigonometric parallax
measurements, or rather uncertain values \citep[e.g.][]{Jeffries95}. 
Previous attempts to calculate
secular parallaxes for young field stars have either applied
the method inconsistently, not propagated uncertainties, or both 
\citep[e.g.][]{Eggen95}.
{\it Can we estimate distances to young field stars and quote believable 
error bars?} 

\begin{figure}[!ht]
\plotfiddle{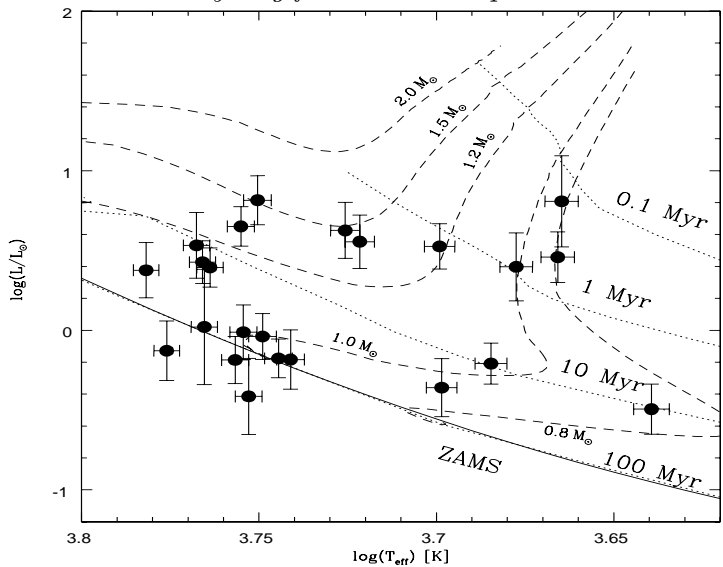}{2.51in}{0}{50}{40}{-175}{-75}
\caption{
HRD for a subsample of FEPS targets which are
statistically more Li-rich than the Pleiades, and whose proper motions and 
radial velocities are consistent with Local Association
membership \citep{Mamajek04phd}. Isochrones are from \citet{D'Antona97}.
\label{fig_hrd}}
\end{figure}

\citet{Mamajek04phd} developed a preliminary iterative method for
estimating secular parallaxes that is applicable to field
stars younger than the Pleiades ($<$125 Myr-old; hereafter
the ``Local Association''). Although the Local Association is
not a cluster in the classical ``coeval'' sense, the kinematics
of field stars that are more Li-rich than the Pleiades 
are remarkably coherent \citep[e.g.][]{Wichmann03}, 
and can be approximated with a simple kinematic model. 
The input parameters are (a) a star's position, (b) proper
motion, (c) radial velocity, and (d) a velocity model with mean 
heliocentric velocity at Sun's position
($U_o, V_o, W_o$), velocity dispersion ($\sigma_U, \sigma_V, \sigma_W$),
and Oort parameters ($A,B,C,K$). The star's parallax is
estimated iteratively using the relation $\pi_{sec}$ 
= ${A\mu}$/${V_{tan}}$,
where $\pi_{sec}$ is the secular parallax, 
$A$ is the astronomical unit (4.74),
$\mu$ is the proper motion, and $V_{tan}$ is the tangential velocity
predicted by the model. One starts with an initial guess distance
to the star (100 pc), calculates the velocity vector 
($U,V,W$) for the velocity
field at the guess distance (3D position), calculates a 
membership probability 
and secular parallax,
and uses the new distance from the secular parallax to revise the
estimate of the velocity vector, etc. These steps are repeated 
until convergence. 
For stars with high membership probability, that are situated
far from the group's convergent point, typically 
$\sim$5 steps are required for the secular parallax
estimates to converge to 1 part in 10$^6$ {\it precision}
(not accuracy!). 
Membership of the star to the kinematic group is tested with
a $\chi^2$ statistic which compares how well the
proper motion direction and measured radial velocity
match that predicted by the velocity field model.
The final membership probability
is used to assess whether the star's motion is consistent with that
predicted by the model, and whether the distance estimate should be
retained.

Using a subsample of two dozen targets in the FEPS Legacy program
which are statistically more Li-rich than the Pleiades (and
hence plausibly $<$125-Myr-old), \citet{Mamajek04phd} calculated
secular parallaxes using an empirical ``Local Association'' velocity
model. Three-quarters of the Li-rich FEPS targets had proper motions
and radial velocities statistically consistent with being Local 
Association members.
The resultant HRD positions and plotted in Fig. 1 and overlaid with
theoretical isochrones. Roughly half of the Li-rich stars are 
consistent with being pre-MS ($<$40 Myr for $\sim$1\,M$_{\odot}$). 
For the few
stars with accurate Hipparcos parallaxes, the agreement between the 
secular parallax distances and Hipparcos trigonometric parallax distances 
is good (e.g. AO Men: 50\,$\pm$\,9 pc vs. 37\,$\pm$\,1 pc; 
HD 202917: 53\,$\pm$\,7 pc vs. 46\,$\pm$\,2 pc). A few objects had
exceptionally young isochronal ages (e.g. HD 285372, RX J1111.7-7620), 
however these appear to be associated with the
Tau or Cha star-forming regions, so their young ages ($\sim$1 Myr)
are not unexpected. A few of the FEPS targets are particularly
close. Among those lacking published parallax estimates, 
\citet{Mamajek04phd} found the following stars to be particularly
nearby and young: RE J0137+18A (64\,$\pm$\,8 pc; $\sim$9 Myr old), 
HD 286264 (71$\pm$\,11 pc; $\sim$22 Myr), and
HD 141943 (67\,$\pm$\,7 pc; $\sim$13 Myr). More results will be
forthcoming in Mamajek (in prep.) and a paper on the
ages of FEPS targets (Hillenbrand et al., in prep.). 

\acknowledgements
EM thanks Michael Meyer and the rest of the FEPS Legacy Science team for useful discussions 
during the course of the author's thesis work. 
EM is supported by a Clay Fellowship through the Smithsonian Astrophysical Observatory (SAO).

\end{document}